\documentclass[10pt]{article}
\usepackage[OME]{express}
\usepackage{booktabs}
\newcommand{\tb}{\textbf}
\usepackage{amsmath,amssymb,graphicx,wrapfig}
\usepackage{indentfirst}
\usepackage{textcomp, tabularx}      
\usepackage{url}
\usepackage{multirow}
\usepackage{array}
\usepackage{todonotes}

\begin{document}

\title{Ar/Cl$_{2}$ etching of GaAs optomechanical microdisks fabricated with positive electroresist}

\author{Rodrigo Benevides$^{1,2}$, Micha\"{e}l M\'{e}nard$^{3}$, Gustavo S. Wiederhecker$^{1,2}$ and Thiago P. Mayer Alegre$^{1,2,\dagger}$ }

\address{
\authormark{1}Applied Physics Department, Gleb Wataghin Physics Institute, University of Campinas, 13083-859 Campinas, SP, Brazil \\
\authormark{2}Photonics Research Center, University of Campinas, Campinas 13083-859, SP, Brazil \\
\authormark{3}Department of Computer Science, Universit\'{e} du Qu\'{e}bec \`{a} Montr\'{e}al, Montr\'{e}al, QC H2X 3Y7, Canada
}

\email{\authormark{$\dagger$}alegre@unicamp.br} 



\begin{abstract}
A method to fabricate GaAs microcavities using only a soft mask with an electrolithographic pattern in an inductively coupled plasma etching is presented. A careful characterization of the fabrication process pinpointing the main routes for a smooth device sidewall is discussed. Using the final recipe, optomechanical microdisk resonators are fabricated. The results show a very high optical quality factors of $Q_\text{opt}>2\times 10^{5}$, among the largest already reported for dry-etching devices. The final devices are also shown to present high mechanical quality factors and an optomechanical vacuum coupling constant of $g_{0}=2\pi\times 13.6$~kHz enabling self-sustainable mechanical oscillations for an optical input power above $1$~mW.
\end{abstract}






\section{Introduction}

Harnessing the confinement of light with wavelength-scale waveguides and cavities has enabled the realization of table-top scale nonlinear optical phenomena in silicon-compatible photonic chips. Recent examples show the versatility of this integrated photonics approach, such as frequency comb generation~\cite{Stern2018,Karpov2018}, quantum computation~\cite{Grafe2016, Sharping2006,Mohanty2017,Qiang2018, Wang2018Science,Faraon2011}, low voltage eletro-optical modulation~\cite{Wang2018}, and optical to microwave coherent conversion~\cite{Balram2016}. One key ingredient that may also boost this revolution is the interaction between light and mechanical degrees of freedom, enabling both read-out and actuation of mesoscale mechanical modes in waveguides~\cite{Wiederhecker2019} and cavities~\cite{Aspelmeyer2014,Riedinger2015,Chan2011,Navarro-Urrios2016}, paving the road towards the sound circuits revolution~\cite{Safavi-Naeini2019,Eggleton2019}. Throughout the quest for better suited cavity geometries that can host optical and mechanical waves, microdisk cavities have proven to be a simple and effective choice. Their tight radial confinement of whispering gallery optical waves and rather strong interaction with radial breathing mechanical modes lead to high optomechanical coupling rates~\cite{Ding2010, Jiang2012b, Sun2012}, which are necessary for efficient device-level functionalities~\cite{Krause2012, Hill:2012cka, Wang2018Science, Metcalfe2014}. Other ingredients in the optomechanical enhancement are the cavity or waveguide material properties. Although silicon is widespread in many optomechanical devices~\cite{EicChaCam0906, Chan2011, Benevides2017b, Santos2017,Luiz2017} due to its mature fabrication, the interest in III-V materials for optomechanical devices has increased~\cite{Xiong2012, Liu2011,Usami2012,Guha:17,Balram2014a} due to their unique optical, electronic and mechanical properties. Gallium Arsenide (GaAs), for instance, is advantageous because of its very high photoelastic coefficient~\cite{Dixon1967}, which leads to large electrostrictive forces and optomechanical coupling~\cite{Ding2010,Baker2014}. Moreover, optically active layers can be easily grown on GaAs wafers, allowing the fabrication of light-emitting optomechanical devices~\cite{Yang2015, Princepe2018, Xi2019}.

Despite the tougher challenges in fabricating high optical quality factor in GaAs cavities, in comparison to the more mature fabrication of silicon, both wet and dry chemistry etching have been successfully developed. Wet etching, followed by surface passivation, resulted in record high intrinsic optical quality factors of $6\times10^6$~ in GaAs microdisks\cite{Guha2016}. Nonetheless, the fabrication of large aspect ratio features and small gaps challenge this route due to wet etch isotropic characteristics and diffusion-limited etching reaction. One alternative is to use inductively coupled plasma reactive ion etching (ICP-RIE)~\cite{Chen2000,Liu2015,Matsutani2016,Faraon2009}, which enables independent control of the plasma density and ion acceleration, allowing a finer control of side-wall roughness and verticality. The difficulty with this technique is the low resist etch resistance that often demands the use of hard etching mask -- in general silicon nitride~\cite{Srinivasan2005}. The need for a hard mask not only complicates the lithography process but also requires a mask removal step.

In this work we report an optimized fabrication process for GaAs-based optomechanical devices that combines standard dry etching chemistry with an electro-resist soft mask, yielding high optical and mechanical quality factors. By avoiding hard masks we ease the GaAs fabrication process and widen its potential exploration in more complex optomechanical cavity designs~\cite{Santos2017}. A complete characterization of the most important ICP-RIE parameters is included, offering a path for for the fabrication of GaAs optical microdevices. By controlling the sidewall roughness and verticality, we produced microdisks with optical quality factor as high as $Q= 2 \times 10^{5}$ -- among the highest reported using soft or hard-mask assisted dry etching~\cite{Srinivasan2005,Faraon2009,Buckley2014, Balram2014a,Rivoire2011}. Moreover, the mechanical characterization of the devices shows mechanical modes with quality factors up to $Q_\text{mec}=760$ and an optomechanical vacuum coupling rate of $g_{0}=2\pi \times 13.6$~kHz.



\section{Microfabrication process}
\label{sec:examples}
\begin{figure}[!b]
\centering
\includegraphics{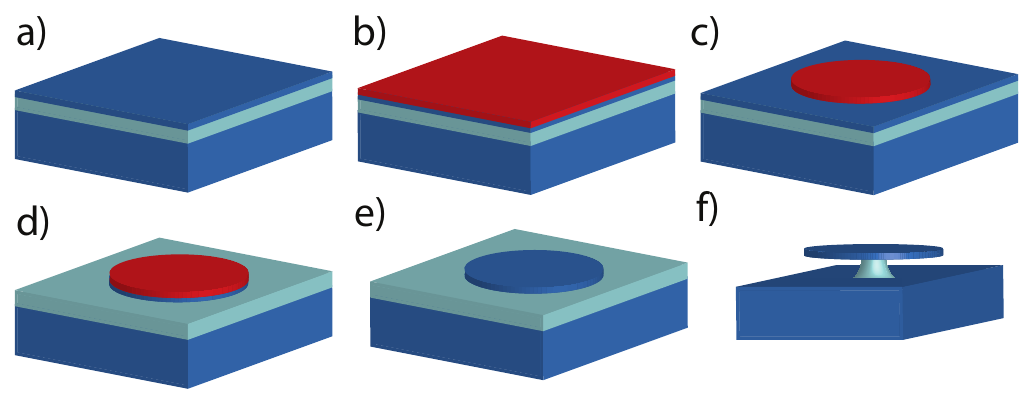}
\caption{\tb{Fabrication procedure.} Over a MBE-epitaxy grown GaAs/AlGaAs wafer (250~nm/2000~nm)(\tb{a}), an electroresist layer (500~nm) is spun (\tb{b}). Electron beam lithography features the resist (\tb{c}), with transfer to GaAs layer done with ICP-etching (\tb{d}). The resist is removed (\tb{e}) and a wet HF-release with post-cleaning is performed, yielding a suspended disk (\tb{f}).}
\label{fig:fabsteps}
\end{figure}
The fabrication steps used in this work follow those of a standard top-down microfabrication process. We start with an intrinsic GaAs/Al$_{0.7}$Ga$_{0.3}$As wafer (250 nm/2000 nm). These wafers were grown  over a GaAs substrate using  molecular beam epitaxy (MBE) (CMC Microsystems, Canada). This technique enables control at the atomic monolayer level~\cite{Madou2002FundamentalsMiniaturization}. It uses an atomic beam of the materials to be deposited in an ultra-high vacuum atmosphere ($\sim 10^{-11}$~Torr) while the substrate is kept at a modest temperature ($400-800^\text{o}$C), when compared to other techniques, which reduces impurity atoms diffusion along the deposited layer. Therefore, a high purity device layer of GaAs is obtained, reducing imperfections in microdevices fabrication that also impact optical and mechanical quality factors of the final devices.

A thin layer of electroresist (ZEP520A - Zeon Chemicals) is spun (Figs.~\ref{fig:fabsteps}\tb{a)-b)}) over the MBE-deposited GaAs layer. Then, the electroresist is patterned with electron-beam lithography, as detailed in section~\ref{sec:elec-dose}, yielding the microdisk shaped patterns illustrated in Fig.~\ref{fig:fabsteps}\tb{c)}. The next step consists of the removal of the superficial GaAs layer using ICP-RIE  chlorine-based plasma etching. Although this is a straightforward method, it possesses a large number of parameters (chamber pressure, gases flow, RF powers) that can be tuned to improve surface roughness and control the sidewall etching angle (anisotropy). In sections \ref{sec:gas-flow} and \ref{sec:ch-pres} we explored several of these parameters in order to understand their role in the etching process. Next, a wet release step of the devices if performed and finalized with a surface cleaning procedure detailed in section \ref{sec:cleaning}.

\begin{figure}[!t]
\centering
\includegraphics[scale=0.8]{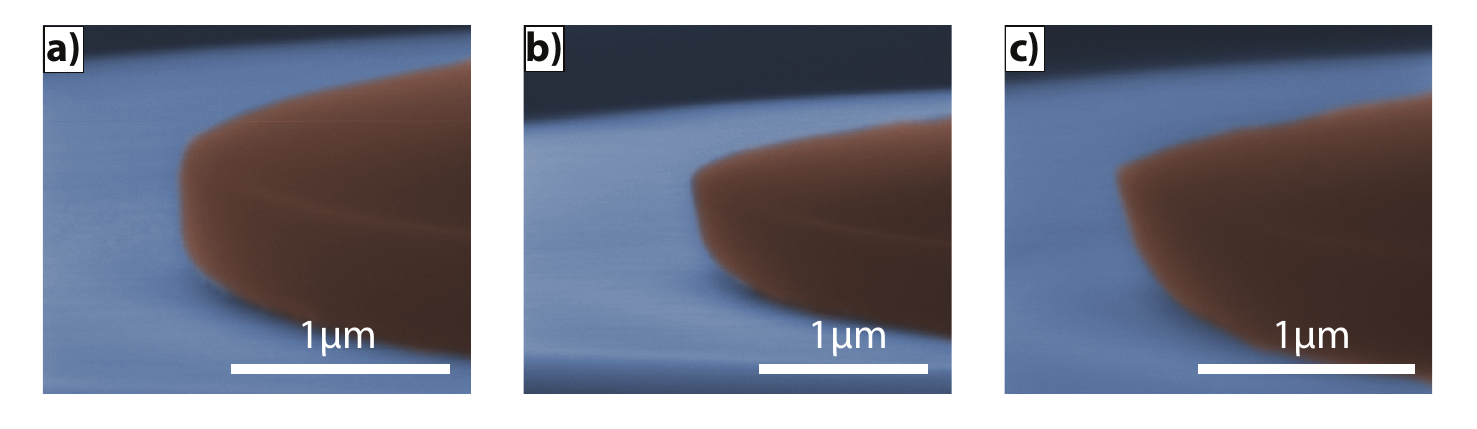}
\caption{\tb{Electroresist edges profile}. Different walls slope are obtained changing the electron beam exposure dose from \tb{a)} 45 \textmu C$/$cm$^{2}$, to \tb{b)} 60 \textmu C$/$cm$^{2}$, to \tb{c)} 75 \textmu C$/$cm$^{2}$. All images were taken at 30 kV using secondary electron detector.  The resist layer was false-colored as red.}
\label{fig:angle}
\end{figure}

\subsection{Lithography parameters}
\label{sec:elec-dose}

The biggest challenge in using an electroresist soft mask for GaAs etching is their low etch resistance. Here we use a positive electro-resist, ZEP520A, which has both high sensitivity and spatial resolution, yet its etch selectivity to GaAs is high enough to  withstand the dry etching process. In this work, we used a $30$~kV electron beam lithography tool (eLINE Plus from Raith Inc.).

The sample were cleaned with hot acetone ($\sim 50^{\circ}$C) and hot isopropyl alcohol ($\sim 70^{\circ}$C), for 5 minutes, followed by a 30 seconds HF:H$_{2}$O (1:10) dip to remove native oxide. The samples were pre-baked for $5$ minutes in a hot plate at $180^{\circ}$C, in order to remove residual water and improve adherence of the electroresist. Then ZEP520A is spun at $2000$ rpm during $60$ seconds, resulting in an $\approx 500$~nm thick resist. Thereafter, we bake the sample again at $180^{\circ}$C during 2 minutes, to evaporate solvents present in the resist.

Electron diffusion through the resist and scattering are known to affect the angle of the developed resist pattern edges.  An exposure dose test was performed to find the optimum doses for our samples, as shown in  Fig.~\ref{fig:angle}\tb{a)}-\tb{c)}. At higher doses, a clearly sloped edge is formed. We choose to work with the vertical profiles as the patterned disks diameters are more accurate.

\subsection{Reflow of electroresist}
\label{sec:reflow}

\begin{figure}[!t]
\centering
\includegraphics[scale=0.85]{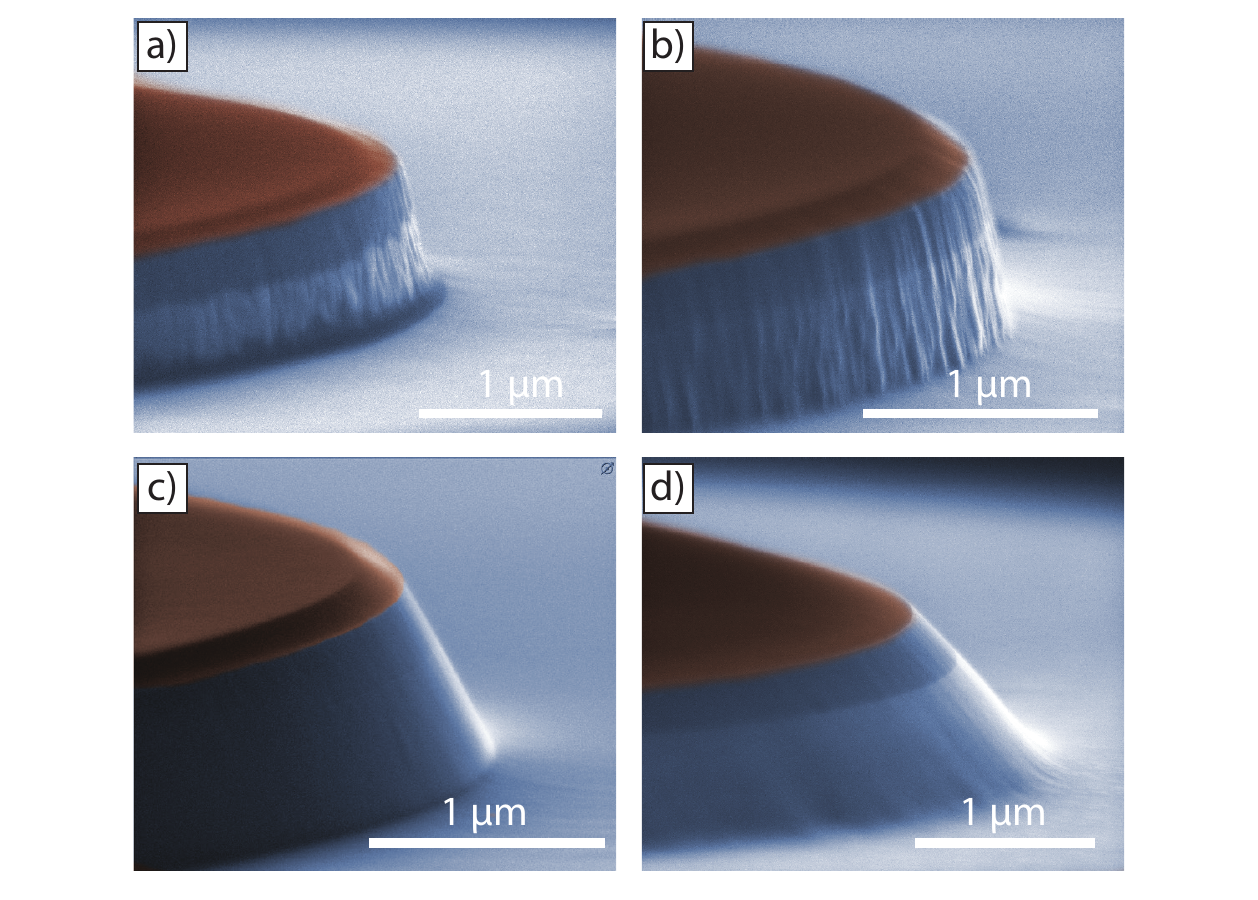}
\caption{\tb{Reflow process}. Microdisks fabricated with resist reflowed at different temperatures: \tb{a)} without reflow, \tb{b)} $140^\text{o}$~C, \tb{c)} $160^\text{o}$~C and \tb{d)} $180^\text{o}$~C for 2 minutes at a hot plate. Etching parameters were gas flow Ar/Cl$_{2}=12/8$~sccm, RF power $=150$~W, ICP power $= 210$~W, and chamber pressure $= 4.5$~mTorr. All images were taken at 20 kV using secondary electron detector. The top resist layer was false-colored as red, the brightness contrast between the top GaAs layer and AlGaAs is intrinsic to the SEM image. The image in \tb{c)} is also shown in Fig.~\ref{fig:ar/cl2}\tb{b)}  and Fig.~\ref{fig:pchamber}\tb{a)}, for easier comparison.}
\label{fig:reflow}
\end{figure}

Minimizing the resist's sidewall roughness is critical as it will be transferred to the GaAs layer during the dry etching process. However, evaluating developed resist sidewall roughness through SEM images is rather challenging. Indeed, despite the apparent sidewall smoothness, show in in  Fig.~\ref{fig:angle}, the etched GaAs sidewalls still presented noticeable roughness after plasma etching, as it is noticeable in Fig.~\ref{fig:reflow}\tb{a)}. This suggests that that further treatment of the developed resist is  required to improve sidewall roughness.

In order to reduce the sidewall roughness transferred to the GaAs we apply thermal reflow of the patterned resist, which is another advantage of using ZEP 520A. This step consists in baking the resist on a hot plate at a temperature higher than the resist softening point, allowing its molecular redistribution and internal stress relaxation that prevent cracks. Previous results with ZEP520A suggest a reflow temperature of $140-145^\text{o}$C~\cite{Kirchner2016}, as suggested by the manufacturer. However, this temperature is not high enough to significantly reduce roughness, as shown in Figs.~\ref{fig:reflow}\tb{a)}-\tb{b)}. This can be attributed to a higher molecular resist weight  after exposure, which has been reported to translate into a higher reflow temperatures for resists based on molecular chain scission~\cite{Kirchner2016,Pfirrmann2016,Schleunitz2014}.  Indeed, an increase of the reflow temperature significantly improves sidewalls profile, as shown in Fig.~\ref{fig:reflow}\tb{c)}-\tb{d}). This confirms our hypotheses that the exposed resist still exhibited residual roughness, despite it was hard to observe it in the SEM images shown in Fig.~\ref{fig:angle}. It is also clear from Fig.~\ref{fig:reflow} that the reflow process impacts the angle of the patterned disk sidewall~\cite{Kirchner2016}. As a consequence,  angled walls are transferred to the GaAs layer.

\subsection{\label{sec:gas-flow} Argon (Ar) and Chlorine (Cl$_2$) flow}

\begin{figure}[!t]
\centering
\includegraphics[scale=0.85]{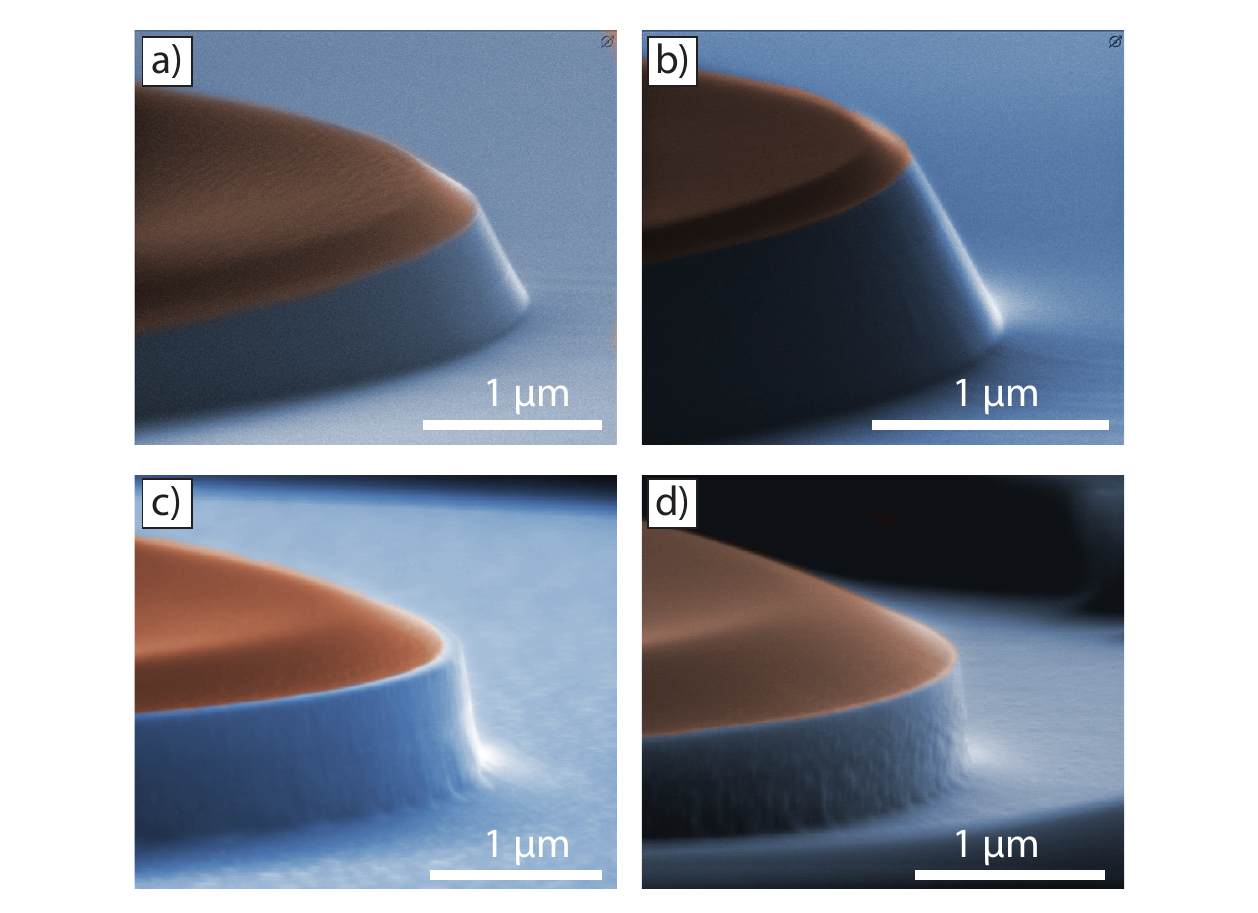}
\caption{\tb{Gases flow}. Etched disks sidewall profile for Ar/Cl$_{2}$ flow of \tb{a)} $16/4$~sccm, \tb{b)} 12/8~sccm, \tb{c)} 8/12~sccm and \tb{d)} 4/16~sccm. Etching parameters were RF power $=150$~W, ICP power $= 210$~W and chamber pressure $= 4.5$~mTorr and a $2$ minutes resist reflow at $160^{\circ}$~C. The etch duration were 90~s for \tb{a)} and \tb{b)}, 35~s for \tb{c)} and $25$~s for \tb{d)}. All images were taken at $20$~ kV using secondary electron detector and false-colored. The image in \tb{b)} is also shown in Fig.~\ref{fig:reflow}\tb{c)}  and Fig.~\ref{fig:pchamber}\tb{a)}, for easier comparison.}
\label{fig:ar/cl2}
\end{figure}

A typical ICP-etching equipment allows control over several etching parameters, including gases flow, chamber pressure and plate and coil powers. In this section, we describe the role of gases flow in GaAs/AlGaAs etching with the ZEP520A masks. We fixed the total gas flow ($20$ sccm), such that changes in the gases proportion would not impact the chamber base pressure. Different proportions of the mixture Ar+Cl$_2$ were used and the profiles obtained can be seen in Fig.~\ref{fig:ar/cl2}. Throughout these steps, we used a reflowed resist mask in the most suitable temperature (2 min at $160^\text{o}$C) in order to isolate  the etch chemistry sidewall roughness. 

Fig.~\ref{fig:ar/cl2}\tb{a)} shows a very smooth wall achieved with a high percentage ($80\%$) of argon in the mixture. However, a bit of roughness is still noticeable  on the resist's upper surface, what could reflect on a higher amount superficial defects in the disks. This is caused by the sputtering resulting from the presence of argon ions in the plasma and can be reduced by increasing the proportion  of chlorine in the mixture, as shown in Fig.~\ref{fig:ar/cl2}\tb{b)}. Further increase of the chlorine flow significantly enhances chemical reactions in the plasma, resulting in a rougher etch, as can be seen in Figs.~\ref{fig:ar/cl2}\tb{c)-d)}. Such roughness sets an upper limit to the chlorine proportion in the mixture.

In addition to the surface roughness of the walls, it is necessary to calibrate the etch rate of each plasma recipe, especially to fabricate devices with tight geometrical constraints~\cite{Santos2017}. These different rates were obtained using an atomic force microscope, which can fully characterize 	the sample topography. The measurements are performed in three steps: the topography of the sample is obtained i) before ICP-etching, ii) after ICP-etching with the electroresist and iii) after removing the electroresist. We can precisely measure the isolated etch rates of GaAs or AlGaAs, and the overall etch selectivity between the resist and GaAs/AlGaAs can be calculated, as shown in table~\ref{tab:gasrate}. We observe a large increase in the GaAs/AlGaAs etching rate when there is more chlorine in the plasma, with a significantly lower resist etching rate. This indicates the possibility of using a plasma of high percentage of chlorine combined with a ZEP520A soft mask to achieve deep-etched~\cite{Mudholkar2007} devices. Due to the large GaAs/AlGaAs etch rate obtained for the highest chlorine flow (4/16 Ar/Cl$_2$), we used shorter duration time for this etch when compared to the other gas mixtures. Contrary to the trend observed for the lower chlorine mixtures, this led to a smaller resist etch rate. Given the etch short duration, this lower rate suggests a non-uniform resist etch rate. 
\begin{table}[!b]
 \centering
 \caption{\bf Etching rate dependence on gas flow.}
  \centering
 \begin{tabular}{c c c}
 \toprule
  Ar/Cl$_2$ flow (sccm) & GaAs/AlGaAs rate (nm/s) & Resist rate (nm/s) \\
 \midrule
	16/4 & 5.9 & 3.7 \\ 
    12/8 & 10.2 & 5.2 \\ 
    8/12 & 22.8 & 8.4  \\ 
    4/16 & 27.2 & 4.8 \\   
 \bottomrule
 \end{tabular}
 \label{tab:gasrate}
 \end{table}

\begin{figure}[!t]
\centering
\includegraphics[scale=0.85]{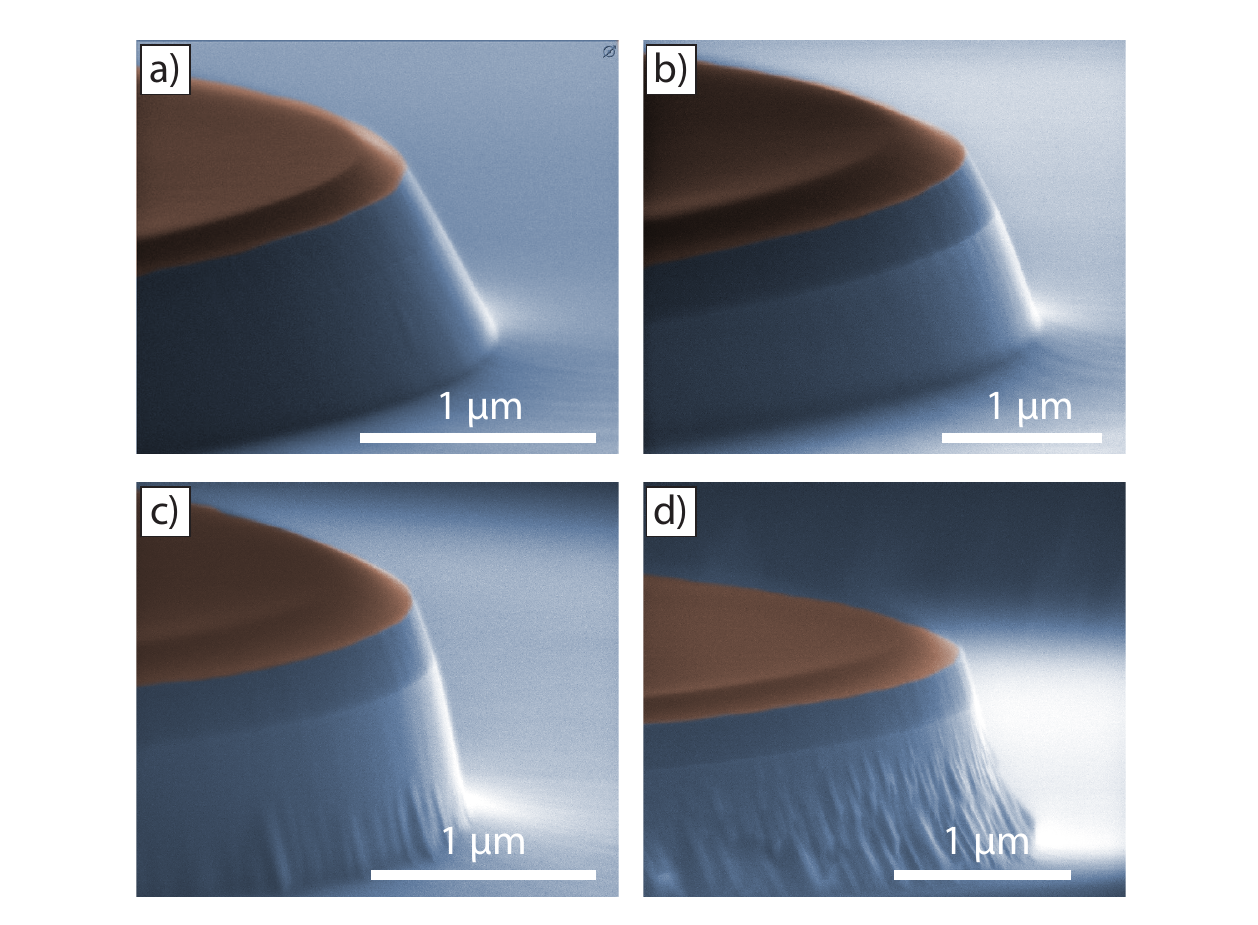}
\caption{\tb{Chamber pressure}. SEM images of microdisks etched at chamber pressures of \tb{a)} $4.5$~mTorr, \tb{a)} $6.0$~mTorr, \tb{a)} $7.5$~mTorr and \tb{a)} $9.0$~mTorr. Changes in sidewall roughness, angle and depth can be observed. These devices were fabricated with Ar/Cl$_2$ flow = $12/8$~sccm, RF power = $150$~W, ICP power = $210$~W and a resist reflow process of $2$ minutes at $160^\text{o}$~C. The image in \tb{a)} is also shown in Fig.~\ref{fig:reflow}\tb{c)}  and Fig.~\ref{fig:ar/cl2}\tb{b)}, for easier comparison.}
\label{fig:pchamber}
\end{figure}

\subsection{Chamber pressure}
\label{sec:ch-pres}
Chamber pressure in ICP etching changes the ions mean free path and can significantly modify the chemical profile of the etch process~\cite{Madou2002FundamentalsMiniaturization}. In order to evaluate the impact of this parameter on the etch results, we fabricated microdisks using different chamber pressures, as can be seen in Fig.~\ref{fig:pchamber}.

\begin{table}[!b]
 \centering
 \caption{\bf Etching rate dependence on  chamber pressure.}
 \begin{tabular}{c c c}
 \toprule
 Pressure (mTorr) & GaAs/AlGaAs  rate (nm/s) & Resist rate (nm/s) \\
 \midrule
	4.5 & 10.2 & 5.2 \\ 
    6.0 & 13.5 & 5.7 \\ 
    7.5 & 16.9 & 3.0  \\ 
    9.0 & 22.5 & 6.0 \\   
 \bottomrule
 \end{tabular}
 \label{tab:shape-functions}
 \end{table}

In general, changes in chamber pressure did not significantly modify sidewall roughness of the GaAs, but it did so for the AlGaAs. An overall selectivity between GaAs/AlGaAs and resist higher than 3.75 was observed, as shown in the table~\ref{tab:shape-functions}, indicating the suitability of this recipe to thicker GaAs-layer devices. We also observe a change in the sidewall angles depending on the pressure, from a $\sim 25^\text{o}$ angle in Fig.~\ref{fig:pchamber}\tb{a)} for $\text{P}_\text{ch}=4.5$~mTorr to an almost vertical wall in Fig.~\ref{fig:pchamber}\tb{d)} for $\text{P}_\text{ch}=9.0$~mTorr. This degree of control of the sidewall angle could be used to manipulate the optical mode overlap with the disk edge, which has been shown to increase the optical quality factors~\cite{Lee2012ChemicallyChip}.

We conclude this section with an optimized fabrication recipe, consisting of an lithography dose of 50~\textmu C/cm$^{2}$, a 2 minutes reflow step at  $ 160^{\circ}$C and ICP-RIE etching  with RF power $=150$~W, ICP power $= 210$~W, Ar/Cl$_2$ flow = $12/8$~sccm and $\text{P}_\text{ch}=4.5$~mTorr. After the etching process, the residual resist is removed by dipping the sample for 5 minutes in trichloroethylene (TCE), 1 minute in acetone, 1 minute in isopropyl alcohol and then blow dry with N$_\text{2}$.

\subsection{\label{sec:cleaning}Undercut  and cleaning steps}
In order to ensure the mechanical degrees of freedom of the microdisk cavities, a  final wet isotropic etching of the AlGaAs layer is performed to undercut the cavity. After the TCE cleaning step, the disks are released using a diluted solution of HF:H$_{2}$O (1:60), which is known to result in AlF$_{3}$ and Al(OH)$_{3}$ residues~\cite{Midolo2015}. These residues are cleaned with a dip in a H$_{2}$O$_{2}$ solution ($30\%$) for 1 minute. A  final dip in a KOH solution ($20\%$ in mass, at $80^{\circ}$C) during 3 minutes is performed to remove oxidized surface layer~\cite{Midolo2015}. These final cleaning steps are crucial to ensure a clean sample and high optical quality factors. 

\section{Optical and mechanical characterization}

\begin{figure}[!b]
\centering
\includegraphics{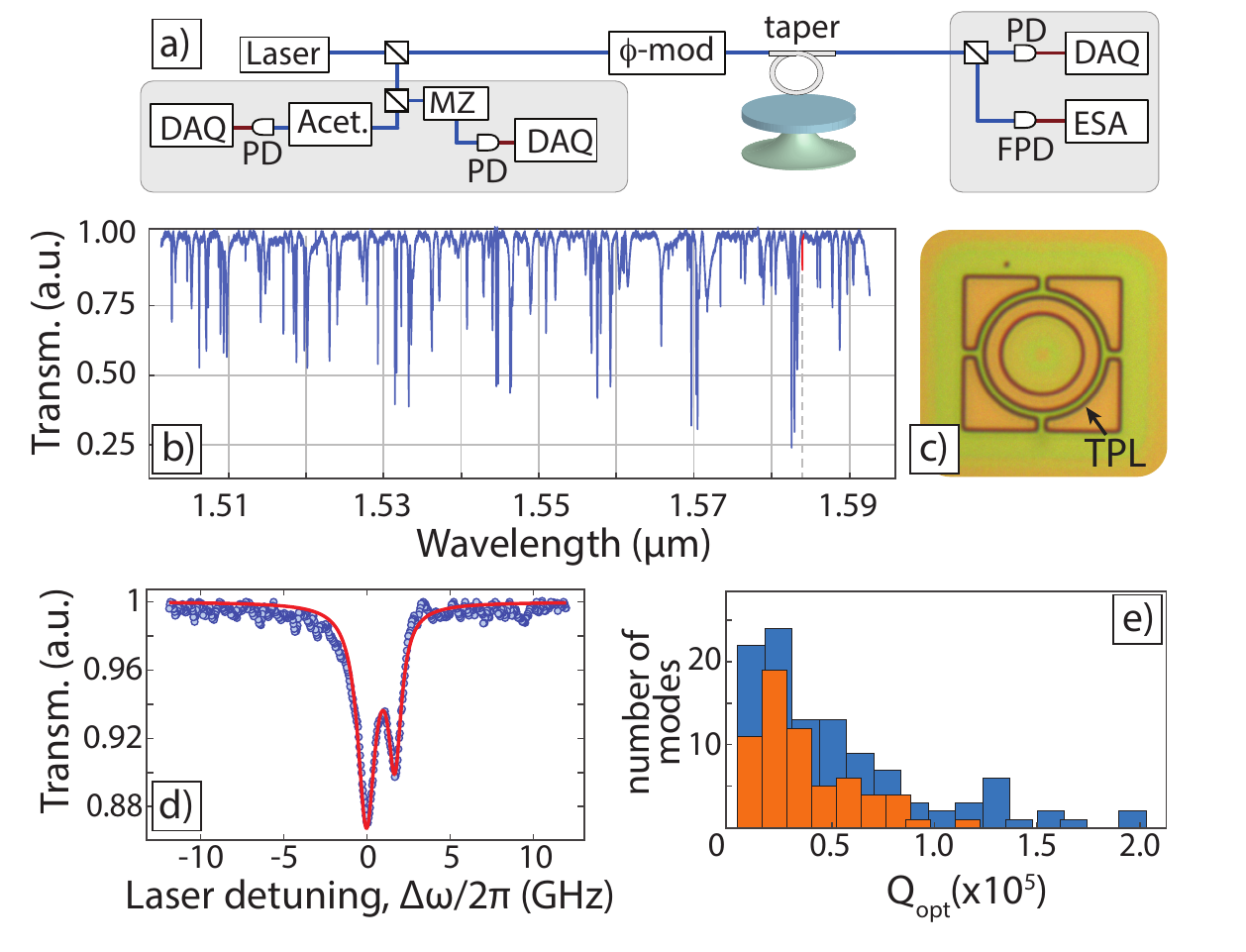}
\caption{\tb{a)} Experimental setup used to characterize optomechanical disks. $\phi$-mod, PD, DAQ, MZ, Acet., FPD and ESA stand for phase modulator, photodetector, analog-digital converter, Mach-Zehnder interferometer, acetylene cell, fast photodetector and electrical spectrum analyzer, respectively. \tb{b)} Broaband spectrum of of a $10~\mu$m radius disk. The fitted intrinsic quality factor is $Q_\text{opt}=2.03\times 10^{5}$. \tb{c)} Microscope image of the fabricated sample, TPL stands for the taper parking lot used to stabilize the tapered fiber position.\tb{d)} Optical modes of a $10~\mu$m radius disk, with intrinsic $Q_\text{opt} = 1.55\times 10^{5}$ and $Q_\text{opt} = 2.03\times 10^{5}$ respectively.  \tb{e)} Instrinsic optical quality factors for a cavity with (blue bars) and a cavity without resist reflow (orange bars). We see that the optimized recipe shows higher quality factors. }
\label{fig:opt}
\end{figure}

The GaAs microdisks fabricated with the optimized recipe exhibit high optical and mechanical quality factors, among the highest reported using soft or hard-mask assisted dry etching~\cite{Srinivasan2005,Faraon2009,Buckley2014, Balram2014a,Rivoire2011}. A schematic of the setup used for device  characterization can be seen in Fig.~\ref{fig:opt}\tb{a)}.  A tunable laser is sent to an optical fiber circuit, an 1\% tap feeds a calibrated fiber-based Mach-Zehnder interferometer and acetylene gas cell for frequency detuning reference and absolute frequency reference. Light is coupled to the cavity through the evanescent field of a tapered optical fiber ($\approx 2$~\textmu m diameter). The transmitted signal is split  (10\%) to feed a transmission monitoring slow photodetector, and (90\%) to feed a fast photodetector ($\approx 800$~MHz bandwidth) to monitor the RF intensity modulation induced on light field by the cavity mechanical modes~\cite{Aspelmeyer2014}. Also, a phase modulator is used to calibrate the optomechanical coupling following the technique by Gorodetsky et al~\cite{Gorodetksy2010}. 

A typical optical transmission spectrum of a 10~\textmu m radius microdisk fabricated with the optimized recipe is shown in Fig.~\ref{fig:opt}\tb{b)} with the highest  optical quality factor resonance highlighted in red. An optical microscopy image of the corresponding device is shown in Fig.~\ref{fig:opt}\tb{c)}. This highest $Q$ mode is also shown in Fig.~\ref{fig:opt}\tb{d)}, where the fitted model reveals an intrinsic optical quality factor of $Q_\text{opt}=2.03\times 10^{5}$; the observed splitting is accounted for in the fitting model and is due to clockwise and counter-clockwise mode coupling. This quality factor is on pair  with those obtained through dry-etching using hard~\cite{Srinivasan2005} or soft mask~\cite{Balram2014a}, yet lower than those of passivated wet-etched disks~\cite{Guha:17}. 

\begin{figure}[!b]
\centering
\includegraphics[scale=1]{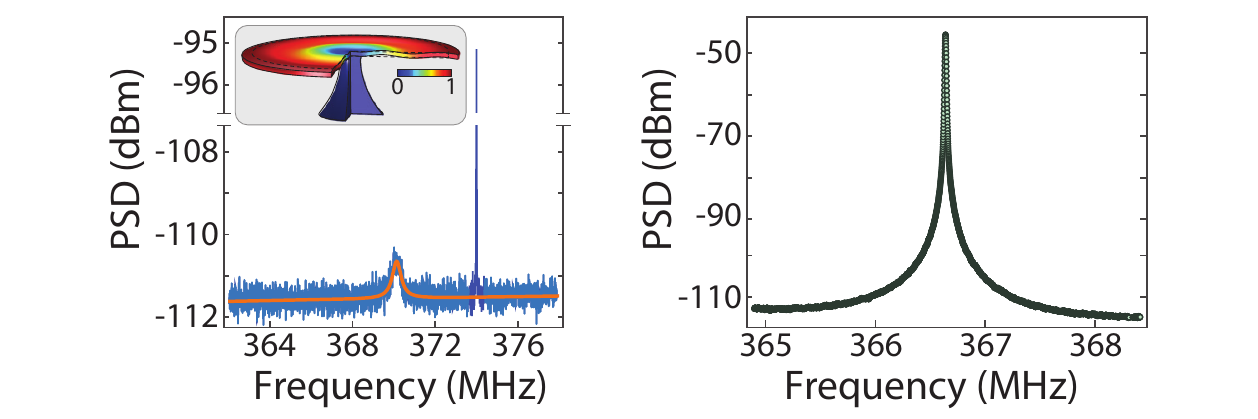}
\caption{ \tb{a)} Mechanical mode at $\Omega_\text{m} = 2\pi\times 370$~MHz observed for a 3.6~\textmu m radius disk, with a quality factor of $Q_\text{mec} = 760$. A phase modulator calibration tone can be seen at $374$~MHz, yielding an optomechanical coupling rate of $g_0=2\pi \times 13.6$~kHz. The inset shows a Finite Element Simulation (FEM) of normalized displacement profile of the fundamental mechanical breathing mode. \tb{b)} Self-sustained oscillation of the fundamental mechanical mode is show with a peak more than $50$~dB above the noise floor.}
\label{fig:mec}
\end{figure}
We have also performed further analysis of the optical modes to infer their transverse mode polarization and radial order. In the  Fig.~\ref{fig:opt}\tb{b)},  it is possible to identify a periodic pattern of the modes, related to the free spectral range (FSR) of the  transverse optical mode families. In fact, we find that the FSR of higher $Q_\text{opt}$ modes closely match those of TE-modes calculated with finite element method (with material dispersion~\cite{Skauli_2003} included). Finally, we have statistically compared our optimized etching recipe but with differing at the resist reflow step. We show the results in Fig.~\ref{fig:opt}\tb{e)} with the quality factors of all modes observed in the tested wavelength range. The orange bars represent the cavity without the resist reflow process, showing a distribution more localized at lower $Q$. On the other hand, the blue bars represent the reflowed device, showing a distribution with higher $Q$  and peak values around $Q_\text{opt}\sim2\times 10^{5}$. This confirms that our optimized fabrication process consistently improved the optical quality of the GaAs microdisks.

To assess the optomechanical performance of our microdisk we analyzed the radio-frequency (RF) spectrum of the transmitted light. For these measurements, a 3.6~\textmu m radius disk was used. Exciting the disk at its highest Q mode at $1563.4$~nm, we were able to observe the RF tone corresponding to the fundamental mechanical breathing mode (see inset of Fig.~\ref{fig:mec}\tb{a)}) at $\Omega_\text{mec} = 2\pi\times 370$~MHz. The corresponding  mechanical quality factor $Q_\text{mec} = 760$ is obtained by fitting the lorentzian shape shown in Fig.~\ref{fig:mec}\tb{a)}. The signal of a calibrated phase-modulator is also shown and was used for the calibration of the optomechanical vacuum coupling constant~\cite{Gorodetksy2010}, yielding $g_0=2\pi\times 13.6$~kHz. This coupling rate is 50\% larger than one would achieve with a similar geometry using a silicon based device, mainly caused by the large photoelastic coefficients of GaAs~\cite{Dixon1967}, responsible for more than 30\% of the total optomechanical coupling rate in this case.

One property of optomechanical devices is that the mechanical modes may experience a back-action force from the optical fields. When the laser is blue-detuned with respect to the mode, the positive feedback force can drive the mechanical system into self-sustained oscillations. Although this is rather straightforward, semiconductor-based  cavities often suffer from detrimental nonlinear loss that prevents reaching this regime. For our devices, when the blue-detuned pump power was increased to  $P \sim 1$~mW, the self-sustained oscillation threshold was easily reached. A strong RF tone ($ > 50$~dBm of signal above noise) was observed and shown in Fig.~\ref{fig:mec}\tb{b)}. It is also possible to see a slight red-shift of the mechanical frequency in the oscillating peak of $\approx 3.5$~MHz, which is attributed to thermal softening of GaAs~\cite{Benevides2018}. This shows that disks produced with our fabrication process can be used as mechanical oscillators, a key aspect to investigate other optomechanical phenomena, such as nonlinear dynamics~\cite{Børkje2013,Krause2015}, coherent optical conversion~\cite{Hill:2012cka} and classical and quantum synchronization~\cite{Zhang2012i, Qiao2018}. Although not explored here, achieving self-sustaining oscillation shows that optical cooling~\cite{Chan2011} is within reach of our devices.

\section{Conclusion}

We have shown that it is possible to control the roughness and the sidewall profile in GaAs microdisks by fully characterizing the ICP plasma etching conditions. This process simplifies fabrication, removing the need of a hard mask and reducing the number of steps in the production of the devices. High optical confinement can be achieved, with optical quality factor as high as $2\times10^{5}$. Further improvement in optical quality factor can be done if one performs surface post-treatment as indicated in~\cite{Guha:17}. Comparing data and simulations, we have evidence that the higher optical quality factor modes correspond to TE-modes, supporting our hypothesis that improvement of the quality factors is due to the reduction in sidewall roughness. We have also investigated the impact of resist reflow in our fabrication process. The observation of radial breathing mechanical modes and excitation of self-sustaining oscillations with low optical powers shows that GaAs devices could also be used in nonlinear optomechanics experiments. 

\section{Funding}
This work was supported by S\~{a}o Paulo Research Foundation (FAPESP) through grants 2016/18308-0, 2012/17610-3, 2012/17765-7, 2018/15580-6, 2018/15577-5 , Coordena\c{c}\~{a}o de Aperfei\c{c}oamento de Pessoal de N\'{i}vel Superior - Brasil (CAPES) (Finance Code 001). The authors thank the CMC Microsystems for providing access to MBE epitaxy and the GaAs wafers and the Center for Semiconductor Components and Nanotechnologies (CCSNano) for the nanofabrication infrastructure.

\tb{Disclosures}: the authors declare no conflicts of interest.


\newpage

\bibliographystyle{osajnl.bst} 
\bibliography{GaAs-OpMEx.bib}

\end{document}